\documentclass[prl,aps,twocolumn,groupedaddress,floats,showpacs,final,superscriptaddress]{revtex4}
\usepackage{graphicx}
\usepackage{bm}
\usepackage{color}
\usepackage{epsfig}
\usepackage{psfrag}
\usepackage{amsmath,amsfonts,amssymb,bm,ulem}

\begin{document}

\author{Igor S. Tupitsyn}
\affiliation{Department of Physics, University of Massachusetts, Amherst, MA 01003, USA}
\affiliation{Russian Research Center ``Kurchatov Institute'', 123182 Moscow, Russia}
\author{Nikolay V. Prokof'ev}
\affiliation{Department of Physics, University of Massachusetts, Amherst, MA 01003, USA}
\affiliation{Russian Research Center ``Kurchatov Institute'', 123182 Moscow, Russia}


\title{Phase diagram topology of the Haldane-Hubbard-Coulomb model}
\date{\today}


\begin{abstract}
We study the phase diagram of the interacting spin-$1/2$ Haldane model with chiral phase
$\phi = \pi/2$ at half-filling. Both on-site and long-range Coulomb repulsive interactions
(Haldane-Hubbard-Coulomb model) are considered. The problem with on-site interaction $U$
alone was addressed in the past by a variety of approximate and finite
size methods that produced results in disagreement with each other both quantitatively and
qualitatively. Here we employ the Diagrammatic Monte Carlo technique to accurately locate
phase transition points to the topologically nontrivial phases in the
$(\Delta, U)$-plane, where $\pm \Delta$ is the inversion symmetry breaking on-site energy,
and establish that momentum dependence of self-energy cannot be neglected in the proper
treatment. We also find that even modest long-range interactions,
typically discarded in theoretical considerations, result in significant shifts of transition
lines.
\end{abstract}


\maketitle

\begin{figure}[tbh]
\includegraphics[width=0.9\columnwidth]{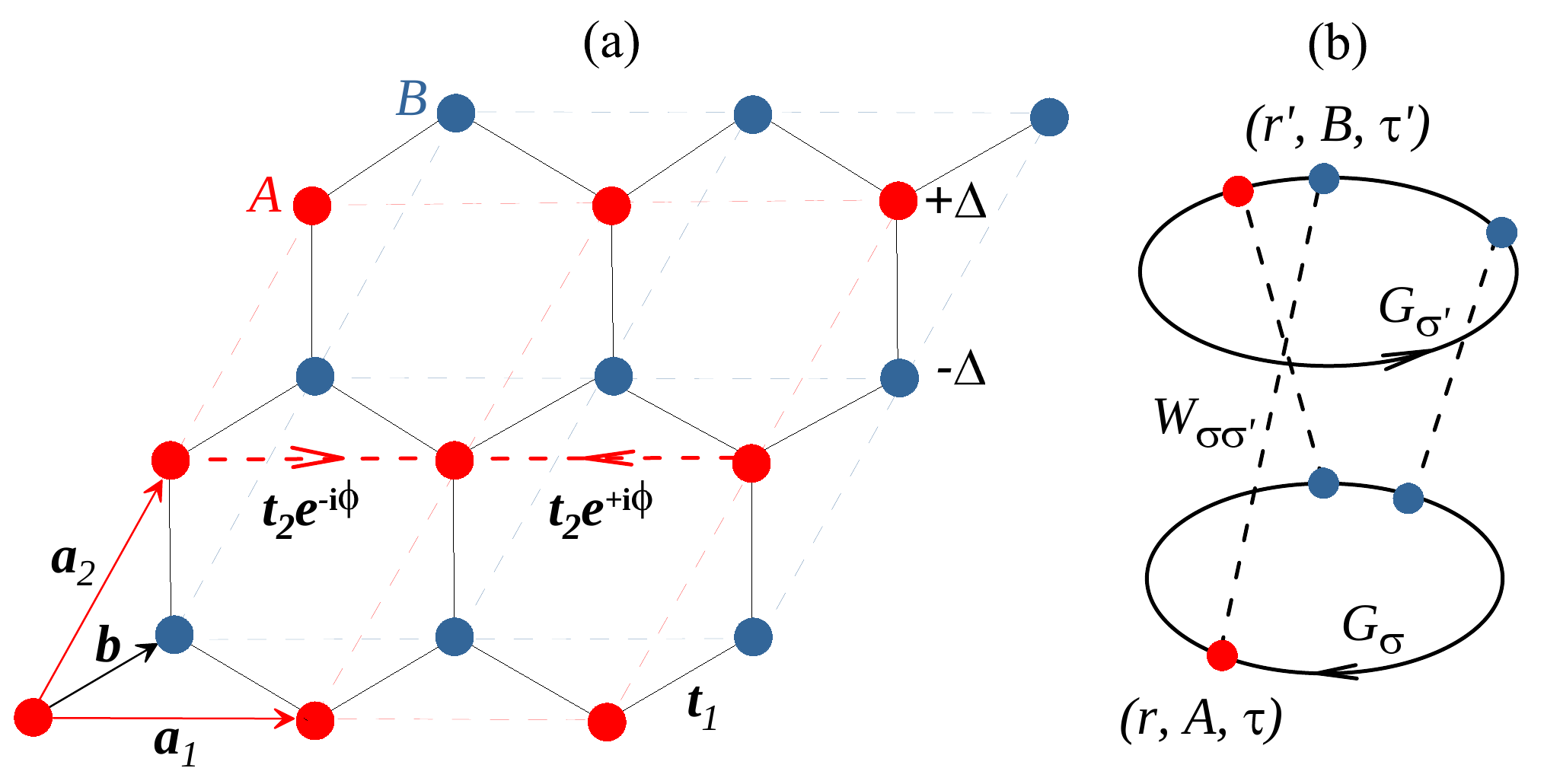}
\caption{ (color online). (a) Decomposition of the honeycomb lattice into
two shifted triangular sub-lattices $A$ and $B$. Lattice spacing $|{\bf a}_1|=|{\bf a}_2|=a$
is used as a unit of length. The n.n. hopping, $t_1$, is real, and the n.n.n. hopping
$t_2 e^{\pm i\phi}$, is complex, with phase $+\phi$ corresponding to counter-clockwise winding
within the hexagon. The staggered on-site energy $\pm \Delta$ has opposite sign on sublattices
$A$ and $B$. (b) Typical $3$-rd order skeleton diagram for free-energy with
${\mathbf r}$ and $\tau$ standing for the unit cell index and
imaginary time, respectively.}
\label{Fig1}
\end{figure}

The Haldane model \cite{Haldane88} was invented to introduce the Integer Quantum Hall Effect
without Landau levels. It describes non-interacting spinless electrons on
the honeycomb lattice with n.n. and n.n.n. hopping amplitudes and inversion symmetry
breaking on-site energy terms $\pm \Delta$, see Fig.~\ref{Fig1}(a).
The n.n. amplitude $t_1$ is real and the n.n.n. amplitude $t_2e^{\pm i\phi}$ is complex,
with chiral phase $\phi$. Complex $t_2e^{\pm i\phi}$ opens a gap at the Dirac points
(the same effect is achieved by non-zero $\Delta$) and breaks the time-reversal symmetry.
The resulting  model features topologically trivial and nontrivial
phases in the $(\Delta, \phi)$-plane, and constitutes the simplest example of a Chern insulator
\cite{ChernIns}.

Its natural generalization to interacting spin-$1/2$ fermions, the Haldane-Hubbard model
(see, for instance, Ref.~\cite{He2011-1}), is considered as one of the key models for studying
topological phases and transitions between them in condensed matter physics.
In recent years it has been intensively studied by various analytical and numerical methods that were
either approximate, such as mean-field (MF) and dynamic mean-field theories (DMFT), or capable of
solving only relatively small system sizes (exact diagonalization), see Ref.~\cite{Troyer2016-1}.
Unfortunately, these calculations produce results that radically disagree with each other
quantitatively, and sometimes even lead to qualitative discrepancies.
For conventional Quantum Monte Carlo methods simulating finite-size systems,
the complex hopping amplitude $t_2$ renders them inefficient due to the notorious
fermionic sign problem.

In general, similarly to the case of the ionic Hubbard model \cite{Dagotto2007}
where $t_2=0$, we expect topologically trivial band and Mott insulator phases
in the limit of large $\Delta$ and $U$, respectively (here $U$ is the strength of on-site repulsion).
In between the two limiting cases, a variety of topologically nontrivial and exotic intermediate
states were proposed (see, for instance, Refs.~\cite{He2011-1,He2011-2,He2012,Hickey2016}).
However, some of these states appear to be ''method specific"; a notable exception is
a topologically nontrivial phase with spontaneously broken spin-rotation $SU(2)$ symmetry
that is found in most mean-field studies \cite{He2011-1,He2012,Huber2016}).
The problem of identifying possible intermediate phases of the Haldane-Hubbard model in the selected
region of parameters, including the one with spontaneously
broken spin-rotational symmetry, has been recently addressed in Ref.~\cite{Troyer2016-1} by three
alternative methods: MF, exact diagonalization (ED), and single-site DMFT.
While all three methods agreed on the identification of possible intermediate phases, they otherwise
demonstrated radical quantitative differences in positions of the corresponding critical points
and lines (see Fig.2 in \cite{Troyer2016-1}).

In this Letter we employ the Bold Diagrammatic Monte Carlo technique (BDMC) developed
for graphene-type systems \cite{GDL2017} to (i) study the phase diagram of the
Haldane-Hubbard model in the same region of parameters as in Ref.~\cite{Troyer2016-1}
and (ii) demonstrate the effect of the often neglected
Coulomb interaction (the corresponding Hamiltonian can be referred to as the Haldane-Hubbard-Coulomb model).
The BDMC technique is not subject to the conventional fermionic sign problem \cite{sign2007,sign2017} and
allows one to deal with arbitrary interaction potential in an approximations free manner \cite{EPI2016}.
The accuracy of final results is controlled by convergence of results with increasing
the expansion order. This approach does work in the most interesting part of the phase diagram
away from the Mott insulating phase.

{\it Model.} The spin-$1/2$ Haldane model on the honeycomb lattice is based on the tight-binding
approximation:
\begin{eqnarray}
H_0 =
&-& t_1 \sum_{<{\mathbf i} {\mathbf j}> \sigma}
(a^{\dag}_{{\mathbf i} \sigma} \;b^{\,}_{{\mathbf j} \sigma} + h.c.) \nonumber \\
&-& t_2 \sum_{<<{\mathbf i}{\mathbf j}>> \sigma}
e^{i \eta_{ij} \phi} (a^{\dag}_{{\mathbf i} \sigma} a^{}_{{\mathbf j} \sigma}
+ b^{\dag}_{{\mathbf i},\sigma} b^{}_{{\mathbf j},\sigma} + h.c.) \nonumber \\
&+& \Delta \sum_{{\mathbf i},\sigma} \xi(i) \; n_{{\mathbf i} \sigma}
- \sum_{{\mathbf i} \sigma} \mu_{\sigma} \; n_{{\mathbf i} \sigma} .
\label{H0}
\end{eqnarray}
The geometry, lattice vectors, and sub-lattice $A-B$ notations are explained in Fig.~\ref{Fig1}(a).
Here $\xi(i \in A)=+1$, $\xi(i \in B)=-1$, and $\mu_{\sigma}$ is the chemical potential for spin component
$\sigma=\uparrow , \downarrow$. The sign of the phase of the n.n.n. hopping amplitude, $\eta_{ij}=\pm$, depends 
on the winding direction, see Fig.~\ref{Fig1}(a). We employ standard second-quantization notations for creation, 
annihilation, and density operators in the site representation for sublattices $A$ and $B$.

In what follows we consider an interacting problem, $H=H_0+H_{\rm int}$, with
\begin{equation}
H_{\rm int} = \frac{1}{2} \sum_{{\mathbf i} {\mathbf j} \sigma \sigma'}
V_{\sigma \sigma'}(\vert {\bf r}_{{\mathbf i}}-{\bf r}_{{\mathbf j}} \vert )
\; n_{{\mathbf i} \sigma} n_{{\mathbf j} \sigma'} \,.
\label{Hint}
\end{equation}
The on-site Hubbard repulsion term $V_{\sigma \sigma'}(0) = U \delta_{\sigma, -\sigma'}$
explicitly takes care of the Pauli principle, while
$V_{\sigma \sigma'}(\vert {\bf r}_{{\mathbf i}}-{\bf r}_{{\mathbf j}} \vert >0 )=
U_{C} |{\bf b}| / \vert {\bf r}_{{\mathbf i}}-{\bf r}_{{\mathbf j}} \vert$ describes the
spin-independent Coulomb tail. Depending on the value of $U_C$, zero {\it vs} non-zero,
Eqs.(\ref{H0}-\ref{Hint}) describe the spin-$1/2$ Haldane-Hubbard or Haldane-Hubbard-Coulomb models.

{\it Formalism}. The BDMC technique employed here is based on stochastic sampling of skeleton diagrams
based on fully dressed Green's functions, $G$, and screened interactions, $W$, or the so-called $G^2W$ skeleton
expansion \cite{Heidin}, see Fig.~\ref{Fig1}(b). At any order of expansion, $N$, self-consistency is reached
by solving Dyson equations that take an algebraic form in the Matsubara frequency-momentum space:
\begin{equation}
G^{-1} = G_0^{-1}-\Sigma \;, \qquad \qquad  W^{-1} = V^{-1}-\Pi\, ,
\label{Dyson}
\end{equation}
where $\Sigma$ is the self-energy and $\Pi$ is the polarization function (both are matrixes in the
spin and sublattice space).
Final results with controlled accuracy are obtained by computing vertex corrections from higher-order
diagrams until convergence is reached. We omit here further technical details as they are fully
documented in Refs.~\cite{KULAGPRL2013,EPI2016}, and, in application to graphene systems, in
Ref.~\cite{GDL2017}.

To obtain the phase diagram in the $(\Delta, U)$-plane we compute the Chern numbers $C^{\sigma}$ and
renormalized electronic dispersions for both spin projections; transitions between topologically
trivial and nontrivial phases manifest themselves by both changing the integer value of $C^{\sigma}$
and by closing the bulk gap at Dirac points. These quantities can be computed by knowing the
fully dressed Green's functions that are the direct outcome of the BDMC simulations.
Following Refs.~\cite{Wang2012,Wang2013}, Chern numbers for an interacting system can be extracted
from properties of the so-called topological Hamiltonian, ${\cal H}_T=-G(i\omega=0,{\bf k})^{-1}$,
assuming that transitions in question are of the ''band-structure" type.
The zero-frequency limit is obtained by extrapolating finite-temperature data
for the set of smallest fermionic Matsubara frequencies, $\omega_n = 2\pi T (n+1/2)$,
with integer $n$ and temperature $T$. Eigenstates of ${\cal H}_T$ then allow
one to compute $C^{\sigma}$ by using the gauge invariant method developed in Ref.~\cite{Fukui2005}.

In the Haldane model $C^{\sigma}$ can take values $0$ and $1$. In what follows we consider the total
Chern number, ${\cal C} = C^{\uparrow} + C^{\downarrow}$,
as a topological order parameter whose allowed values $0, \; 1,$ and $2$ distinguish phases.
Our calculations are performed at half-filling for system sizes $L^2 = 16^2$ and $32^2$ (the number of
sites is $2L^2$) with periodic boundary conditions and at temperatures $T/t_1=0.1$ and $0.05$, to
quantify finite-size and finite-temperature effects. Chern numbers calculated for our system parameters
using the method of Ref.~\cite{Fukui2005} are integer with accuracy better than $10^{-8}$. We take
$t_1=1$ as the unit of energy and fix $t_2=0.2$ and $\phi = \pi/2$, as in Ref.~\cite{Troyer2016-1}.
We had to limit our analysis to on-site repulsion $U \leq 7$; obtaining converged answers at larger
values of $U$ requires reformulation of the diagrammatic expansion and goes beyond the scope of present
work.


{\it Haldane-Hubbard model.} We first study the phase diagram of the Haldane-Hubbard model
(\ref{H0}-\ref{Hint}) when $U_c=0$, and concentrate on the topologically nontrivial
intermediate Chern insulator states away from the Mott insulator.
To obtain transition lines separating the band and Chern insulators we fix
$U$ and find where the total Chern number changes its integer value along the
$\Delta$-axis. If we only account for the first-order diagrams,
equivalent to the so-called fully self-consistent GW approximation, then we do not see
the topologically nontrivial phase ${\cal C}=1$ with spontaneously broken
spin-rotational symmetry. Next-order vertex corrections do not change this
outcome either; i.e., at the level of two lowest orders the skeleton diagrams
results are consistent with the DMFT calculations, but plainly contradict
the MF and ED predictions \cite{Troyer2016-1}. This is a clear sign that
precise location of the point where all three phases meet cannot be determined
reliably by approximate methods.

\begin{figure}[tbh]
\includegraphics[width=0.75\columnwidth]{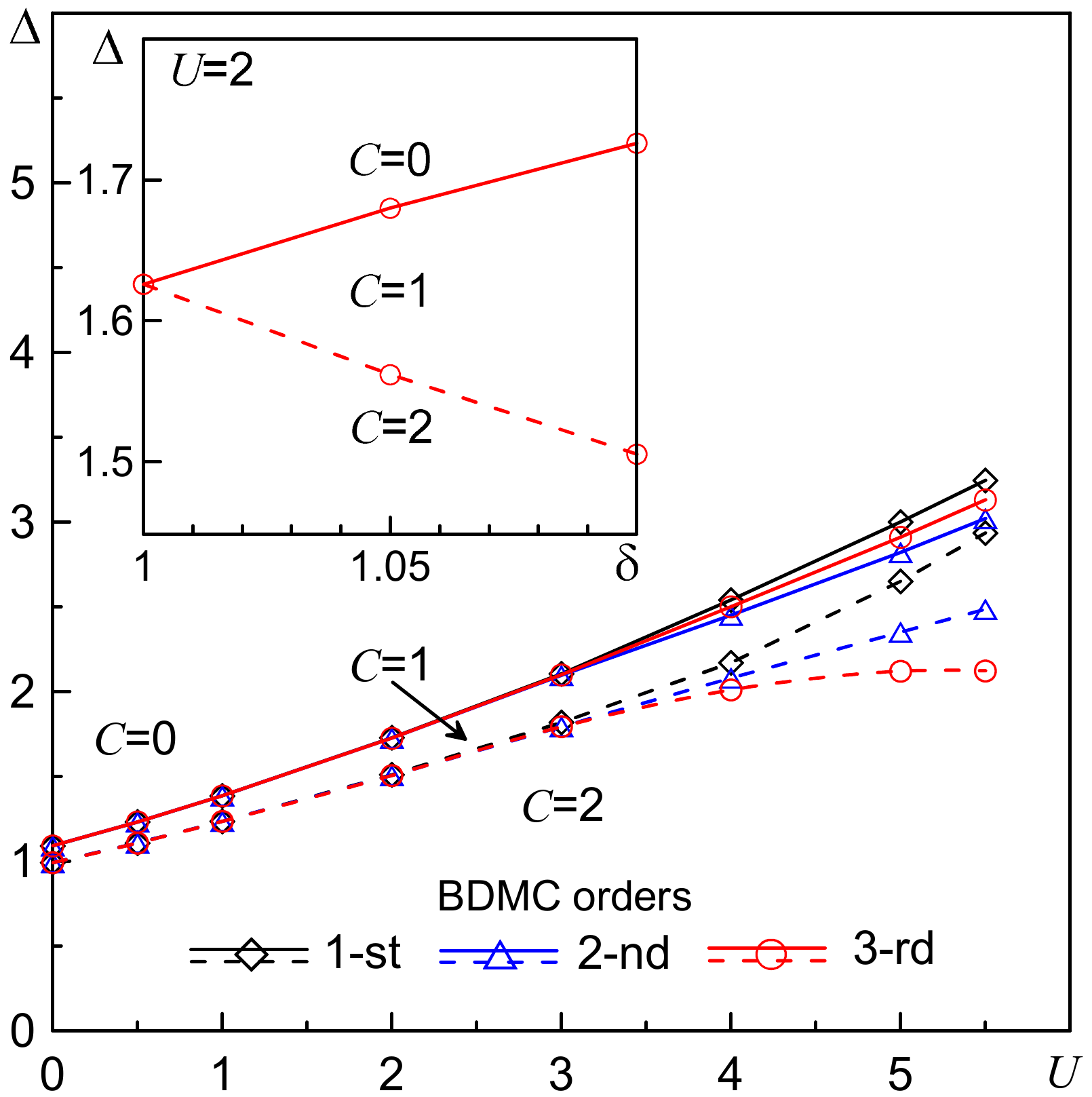}
\caption{ (color online). Chern insulator phases (${\cal C}=2, \; 1$) of the Haldane-Hubbard
model with explicitly broken spin rotational symmetry ($\delta = 1.1$, see text)
in different skeleton orders (${\cal C}=0$ corresponds to the topologically trivial band
insulator phase). In the inset we show how the size of the ${\cal C}=1$ phase for $U=2$
shrinks with the value of spin-imbalance parameter $\delta$. Statistical and systematic
errors in this and other figures are smaller than symbol sizes.}
\label{Fig2}
\end{figure}

To locate the ${\cal C}=1$ phase and eliminate the first-order transition scenarios
we employ the following strategy. In one set of simulations we break the spin-rotational
symmetry explicitly by making the hopping amplitudes spin-dependent:
\[
t_{1,2}(\uparrow)   \to t_{1,2}(\uparrow) / \delta^{1/2}, \;\;
t_{1,2}(\downarrow) \to t_{1,2}(\downarrow) \delta^{1/2}, \; \mbox{with}\;\delta > 1.
\]
In this case, the ${\cal C}=1$ phase exists even at $U=0$, but for $U>4$ converged answers
are obtained only by accounting for high-order diagrams (up to 5-th order), since the
behavior at $N=2$ and $N=3$ is different, see Fig.~\ref{Fig2}. We then use the solutions
for $G$, $\Sigma$, and $\Pi$ to initialize calculations with smaller spin-imbalance all the
way to $\delta=1$ (no spin imbalance) to see if the ${\cal C}=1$ phase survives. We follow
this protocol for all values of $U \leq 5.5$. In the second set of simulations we start with
$\delta=1$ and monitor how results change with increasing $N$, in particular, how the
${\cal C}=1$ state appears in some region of parameters and remains stable. The second protocol
is applied at $U \geq 4$.

Following the first protocol, we determine that the ${\cal C}=1$ phase goes away as
$\delta \to 1$ for all values of $U < 5$, see a typical data set for $U=2$ in the
inset of Fig.~\ref{Fig2}. This rules out the phase diagram topology predicted by the
ED studies of small clusters \cite{Troyer2016-1} (apparently, the momentum space resolution
was too sparse to conclusively eliminate the ${\cal C}=1$ state in this parameter regime).
In the second ($\delta=1$) protocol, the ${\cal C}=1$ phase opens up only in simulations performed
at $U > 5$ with $N \ge 3$.

\begin{figure}[tbh]
\includegraphics[width=0.75\columnwidth]{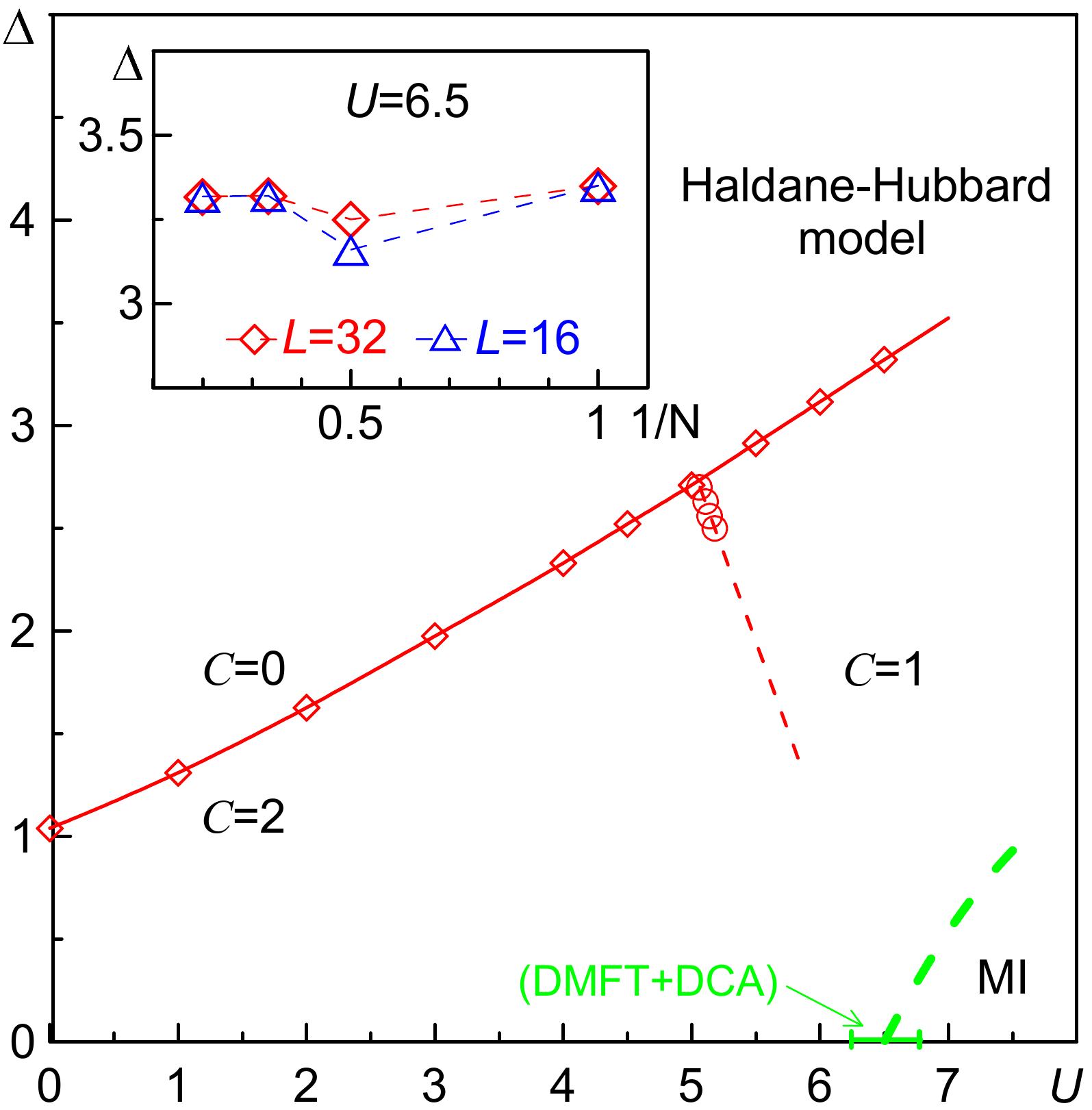}
\caption{ (color online). Phase diagram of the SU(2)-symmetric Haldane-Hubbard model.
Solid red line with diamonds separates the topologically trivial band (${\cal C}=0$) from
Chern insulator phases. Dashed red line with circles separates Chern insultors
with preserved (${\cal C}=2$) and spontaneously broken (${\cal C}=1$) spin rotational symmetries.
Dashed line is extrapolated towards the first-order transition between the
${\cal C}=1$ and Mott insulator phases shown by green dashed line as established in
Refs.~\cite{Troyer2016-1,Troyer2016-2}. In the inset we show how the position of critical
line at $U=6.5$ depends on the inverse skeleton expansion order $1/N$. Within chosen accuracy
of 0.1 we see no difference in converged answers for linear system sizes $L=32$ and $16$ (as well
as for temperatures $T=0.1$ and $0.05$).}
\label{Fig3}
\end{figure}

Results obtained within both protocols are summarized in Fig.\ref{Fig3}. The transition
line, separating the band and Chern insulators, is rather close to the one obtained in
Ref.~\cite{Troyer2016-1} within the single-site DMFT. The ${\cal C}=1$ phase does exist,
but the critical on-site repulsion, $U_m$, where this phase first emerges and the two
transition lines meet is found to be close to $U_m \approx 5$. This value is nearly three
times(!) smaller than the single-site DMFT result for $U_m$, indicating that momentum dependence
of self-energy plays important role in the quantitative analysis. The MF prediction
$U_m \sim 4$ happens to be closer to the correct answer, but the slope of
$({\cal C}=2)$-$({\cal C}=1)$ line has an opposite sign. If we extrapolate our results for the
transition line between the ${\cal C}=2$ and ${\cal C}=1$ phases towards larger values
of $U$ we hit the first-order transition to the Mott insulator state as determined in
Refs.~\cite{Galitski2010,Troyer2016-2}. In other words, our result is consistent with having
only one transition point along the $U$-axis at $\Delta=0$. Unfortunately, the $G^2W$ skeleton
expansion implemented here does not work in the vicinity of the Mott state.

\begin{figure}[tbh]
\includegraphics[width=0.75\columnwidth]{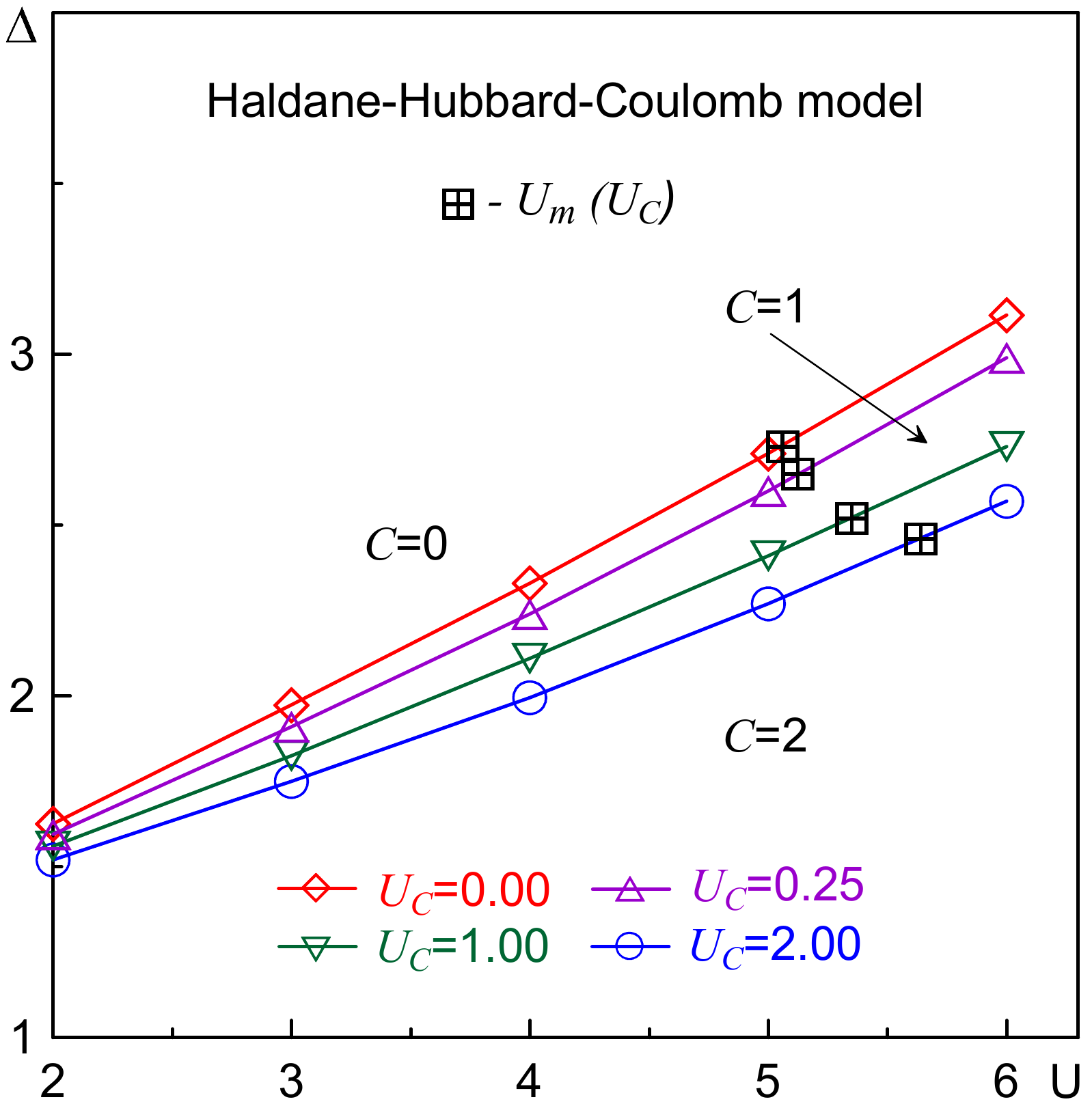}
\caption{ (color online). Effect of the long-range Coulomb potential, $V_{\sigma \sigma'}(r>0)$,
on the phase diagram. Solid lines with symbols separate the topologically trivial band and nontrivial
Chern insulators for different values of the Coulomb coupling $U_C$. Black squares with crosses mark
positions of critical points $U_m(U_C)$, separating the Chern phases with ${\cal C}=2$ and ${\cal C}=1$.}
\label{Fig4}
\end{figure}

{\it Haldane-Hubbard-Coulomb model.} We now proceed with the study of long-range interaction effects
and consider non-zero values of $U_C$ in (\ref{Hint}). To ensure that the repulsive potential
is monotonously decreasing with distance we take $U_C \leq U$ (by definition, $U_C$ is
the strength of the n.n. interaction).
The most obvious effect of the Coulomb potential can be understood as follows.
Imagine that we add a constant interaction term $V_C$ at all distances
$|{\bf r}_i - {\bf r}_j| >0$ (i.e., an infinite-range potential) to the Haldane-Hubbard model.
This would be equivalent to simply shifting the chemical potential of the model by $V_C$ and reducing
the value of the on-site repulsion to $U-V_C$. Correspondingly, under this transformation the entire
solid line is translated horizontally, $\Delta(U,V_C)=\Delta(U-V_C,0)$, and thus appears shifted
downwards in the $(\Delta,U)$ plane, as in Fig.~\ref{Fig4}. However, this thinking is only valid
qualitatively; the horizontal-shift transformation strongly overestimates the downwards shift
and fails to explain the correct locations of special points $U_m(V_C)$ (squares with crosses
do not form a horizontal line).

Overall, Coulomb interactions suppress the ${\cal C}=1$ phase and push it to higher values of $U$ and
lower values of $\Delta$. Given that in realistic materials the ratio between the $U$ and $V_C$ parameters
is not small, Coulomb effects cannot be neglected or easily (as in the above example with constant shift
at $r > 0$) accounted for in quantitatively accurate predictions.

{\it Conclusions.}
We investigated the phase diagram of the spin-$1/2$ Haldane model on honeycomb lattice with on-site
and long-range Coulomb interactions by the Bold Diagrammatic Monte Carlo method to
obtains results with controlled accuracy for convergent skeleton sequences.
We confirmed the existence of topologically nontrivial intermediate phase with spontaneously
broken spin-rotation $SU(2)$ symmetry, where the Chern numbers for two spin components are
$0$ and $1$, resulting in the total Chern number ${\cal C}=1$. This phase emerges only after
we account for vertex corrections beyond the second $G^2W$ skeleton expansion, indicating that
any approximate theoretical scheme would be prone to large quantitative errors. Indeed, for
the Haldane-Hubbard model we found that the transition between the band insulator, ${\cal C}=0$,
and ${\cal C}=1$ phases takes place at $U = U_m \sim 5$, nearly a factor three smaller than the
$U_m$ value predicted by the single-site dynamic mean-field  theory \cite{Troyer2016-1}, which
neglects the momentum dependence of the self-energy. The coarse-grained structure of the obtained
phase diagram is close to that revealed by exact diagonalization \cite{Troyer2016-1} except for
artifacts of momentum quantization in small clusters that prevent one from observing a direct
${\cal C}=0 \longleftrightarrow {\cal C}=2$ transition.

In the case of the Haldane-Hubbard-Coulomb model we quantified effects of typically neglected
long-range Coulomb interactions. Both topologically nontrivial phases survive, but the $1/r$
potential tends to suppress topological phases in favor of the band insulator one and shifts the
${\cal C}=1$ phase towards larger values of on-site repulsion. While remaining quantitative,
Coulomb effects cannot be neglected if one aims at accurate predicting for real materials.

This is the first application of the BDMC technique to properties of interacting topological
insulators. Given that it is applicable to both doped and undoped systems with arbitrary dispersion
relation and shape of interaction potential, in future work it would be interesting to study the
Haldane-Hubbard-Coulomb model at other filling factors and values of $\phi$, and explore cases with
``flat band" dispersion relevant to the search for Fractional Chern Insulator states (Fractional
Quantum Hall Effect without Landau levels) \cite{FracChern}. Our technique is directly applicable to
these type of problems \cite{KULAGPRL2013}.

{\it Acknowledgements.} We thank T. Sedrakyan for discussions. This work was supported by
the Simons Collaboration on the Many Electron Problem, the National Science Foundation under
the grant PHY-1314735, and the MURI Program ``New Quantum Phases of Matter" from AFOSR.

\end{document}